\documentstyle[aps,epsf]{revtex}
\def \bi{\bibitem}

\def \d {{\rm d}}

\begin{document}
\title{How glasses explore configuration space}
\author{ S. Franz\\
}
\address{
The Abdus Salam International Center for Theoretical Physics\\
Strada Costiera 11,
P.O. Box 563,
34100 Trieste (Italy)\\
}
\date{\today}
\maketitle

\begin{abstract}

We review a statistical picture of the glassy state derived from the 
analysis of the off-equilibrium fluctuation-dissipation relations. We define an
ultra-long time limit where ``one time quantities'' are close to
equilibrium while response and correlation can still display aging. 

In this limit it is possible to relate the fluctuation-response
relation to static breaking of ergodicity. The resulting picture
suggests that even far from that limit, the fluctuation-dissipation
ratio relates to the rate of growth of the configurational entropy
with free-energy density.

\end{abstract}

\section{Introduction}
Glassy systems spend long times out of equilibrium, i.e. in regions of
configuration space which have vanishing Boltzmann probability. The
evolution towards more and more optimized regions of phase space is so slow,
that the fast degrees of freedom react on short time scales as if they
were in equilibrium in the frozen background of the slow variables. 
On longer time scales however the off-equilibrium nature of the glassy
phase shows up in the phenomenon of aging: the response to small
perturbations, as well as the correlation functions, depend on the ``age'' of
the glass, i.e. the time spent in the low temperature phase
\cite{struik,aging}.  The
dynamics becomes slower and slower as the age becomes larger and
nevertheless, even on the largest scales observed there is no tendency to
stop: the system eventually wanders away from
any finite region of phase space. 

In the last years, following some developments of spin glass mean field
theory \cite{cuku,frame,cukusk}, many theoretical \cite{francesi}, numerical
\cite{x(q)-sim} and more recently experimental papers \cite{israel,cili,ocio}
have focused on the study of off-equilibrium fluctuation-dissipation relations.
In \cite{cukupe}, it has been shown that the fluctuation-dissipation
relation found in mean-field analysis of glassy relaxation can be used
to define effective temperatures, higher then the one of the heat bath
governing the heat exchanges between slow degrees of freedom.

The analysis of the fluctuation-dissipation relation has revealed deep
relations between aging dynamics and the nature of the free-energy landscape
\cite{fmpp}. In this paper we would like to review a picture of glassy
dynamics based on the analysis of these relations \cite{fv}.

Two kinds of glassy systems emerge from experimental and theoretical
analysis. 
One can identify two kind of glassy behaviors 
\begin{itemize}
\item
A first class, of which spin glasses are an example, where the
asymptotic value of extensive quantities, like internal
energy, and magnetization seem not to depend on the cooling rate or
other differences in the cooling procedure.

\item 
A second class, the one of structural glasses, in which the 
apparent asymptotic value of these quantities depends strongly on the
cooling rate, and remain strongly different from the equilibrium
values for the the largest time scales which can be probed. 
\end{itemize}

Aging phenomena are common to both families. During aging the linear
response to external perturbations displays age dependent behavior
and anomalies with respect to equilibrium, which is conform to the
fluctuation-dissipation theorem. Many efforts have been devoted to
relate these anomalies to the properties of the underlying energy
-or free-energy- landscapes \cite{KuLa,fmpp,fv,BiKu,bkvs,biroli}.

A first step towards the comprehension of this relation can be
achieved idealizing the first situation assuming that 
for long times
{\it all} extensive ``one time observable'' (1TO), or more precisely the
ones that can be written as sum of local quantities, are close to
equilibrium. We call this situation {\it ultra-long time limit} and
observe that in principle a) this limit does not imply the absence of
aging in two time observables such as responses and correlation
functions, b) usual nucleation arguments imply that in systems with
regular short range interactions this limit is achieved in a finite,
i.e. volume independent, time. Of course this last observation remains
a matter of principle in structural glasses, where this equilibration
time is exceedingly long, and one needs to understand dynamics on much
shorter time scale.

For notational simplicity, in the theory we are going to expose, we
will use the language of magnetic systems. Our considerations however
will very general and can immediately applied e.g. to glass forming
systems of classical particles in interaction.  In order to illustrate
the ideas we discuss a spin glass system where the spins $S_x$
interact through a Hamiltonian
\begin{equation}
H(\bf{S})= -\sum_{x<y} J_{x,y}S_x S_y.
\end{equation}
on the D-dimensional square lattice and short range interactions
$J_{x,y}$.  The spins will be for simplicity assumed to be Ising
variables $S_x=\pm 1$. We will consider the dynamical setting of a
fast quench from high to low temperature at an instant marking the
origin of the time axes.

\section{Aging in the ultra-long time limit}

The classical aging experiments concentrate on the study of the
response of the system to an external perturbation in the linear
regime. Consider the effect of an external magnetic field $h$,
corresponding to a perturbation term in the Hamiltonian
$H^{(1)}(\bf{S})= -h \sum_{x} S_x$, acting from the quenching time $0$
to a waiting time $t_w$. At the linear order in $h$, the magnetization
$M(t,t_w)=\frac{1}{N} \sum_x \langle S_x(t)\rangle_h$ at later times
$t$ can be written as $M(t,t_w)=h\chi(t,t_w)$, where, introducing the
instantaneous response $R(t,s)=\frac{\delta\langle M(t)\rangle}{\delta
h(s)}$, the susceptibility can be written as $\chi(t,t_w)=\int_0^{t_w}\;ds\;
R(t,s)$. This is a quantity usually measured in aging experiments and
exhibit scaling, non-time translation invariant behavior. In usual
equilibrium regimes, this function is related to the correlation
function $C(t,t_w)=\frac{1}{N} \sum_{x,y}\langle
S_x(t)S_y(t_w)\rangle$ by the fluctuation-dissipation relation
$\chi(t-t_w)=\beta C(t-t_w)$. Off-equilibrium the same relation has in
general no reason to be valid.  One usually defines the
fluctuation-dissipation ratio (FDR)
\begin{equation} 
\beta \tilde{x}(t,t_w)=\frac{\partial \chi(t,t_w)/\partial t_w} {\partial
C(t,t_w)/\partial t_w}
\end{equation}
or $x(t,C)$ defined by the relation $x(t,C(t,t_w))=\tilde{x}(t,t_w)$.
In equilibrium conditions the FDR is equal to unity and
$\chi(t,t_w)=\beta C(t-t_w)$.  Spin glass mean field theory
\cite{cuku,frame,cukusk}, has suggested that in aging systems $x(t,C)$
tends to a non trivial limit $x(C)$ taking the limit $t,s\to\infty$
for fixed values of $C(t,s)$. In these systems the FDR appears to have
some universal character. It is independent of the particular
response/correlation function probed. The function $x(C)$ appears to
have mathematically the properties of a probability function. The
ratio $T/x(C)$ can be considered an effective temperature governing
heat exchanges on degrees of freedom reacting on the time scales
specified by the value of the correlation \cite{cukupe}. Such non
generic behavior can be understood in the ultra-long time limit,
resulting in a rather detailed statistical description on how glassy
systems visit configurational space in aging dynamics.  In the limit
in which the FDR becomes time independent, one time observables
(e.g. energy and magnetization) become asymptotically close to their
final value.  We will call this limit ``ultra-long time limit'' in
reference to the fact that structural glasses are always observed on
much shorter time scales.  Let us stress that in short range systems
with regular interactions the asymptotic values of the 1TO must be the
equilibrium one. The ultra-long time limit while it could be
appropriate in spin glasses where the values of 1TO's seem  not to  depend on
cooling procedure, but it is certainly not an appropriate description for
structural glasses, where the 1TO are strongly out of equilibrium. In
this context we can still think of this limit as a useful
conceptualization to study aging in a simplified situation and getting
hints on the statistical principles that govern glassy dynamics even
far from this ideal situation. 

\section{Equilibrium ergodicity breaking}

We would like to discuss how a non trivial FDR in the ultra-long time
limit relates to the phenomenon of ergodicity breaking in the
equilibrium distribution. Breaking of ergodicity, or absence of
ergodicity at equilibrium means that the weight of the
equilibrium distribution is concentrated on more then one disjoint
phase space regions, called ergodic component, where the ergodic
property holds separately.  If the system is prepared in one of these
regions it will never get out from it. A familiar example is provided by
systems in presence of a first order phase transition, where different
phases are equally dynamically stable. In that case suitable
``external field'' can be used to project on the different phases. The
situation we would like to describe, that of an ideal glass, can be
expected to be different. In disordered systems, the possible
different ergodic components dominating at low temperature can be
expected to be as disordered as the configurations of the high
temperature phase and no external field is available to select any of
them. A suitable description of that situation has been provided by
spin glass mean field theory \cite{MPV} where ergodicity breaking 
is characterized  statistically through the comparison of the different
ergodic components. One first define a measure of similarity
among configuration, and then studies its statistical distribution
induced by the canonical measure and the eventual quenched
disorder. For spin glasses it is natural to define the overlap between
two spin configurations ${\bf S}=(S_1,...,S_N)$ and ${\bf
S'}=(S'_1,...,S'_N)$ as the normalized scalar product:
\begin{equation}
q({\bf S},{\bf S'})=\frac 1 N \sum_i S_i S'_i
\end{equation}
The overlap probability function (OPF), is defined as 
\begin{equation}
P(q)=\overline{ \frac{1}{Z^2} \sum_{{\bf S},{\bf S'}}\exp\big(-\beta
(H({\bf S})+H({\bf S'}))\big)\delta (q-q({\bf S},{\bf S'}))}. 
\end{equation}
where the over-line denotes the average over the quenched variables of
the system (if any).  If ergodicity holds the OPF equal to a single
delta function. Any difference from this simple form is sign of
ergodicity breaking. Conversely, a single delta function is not
necessarily associated to ergodicity.  The different components
$\alpha=1,2,3,...$ appear in the equilibrium distribution each with a
probability weight $w_\alpha$ reflecting its free-energy difference
with the lowest state. It turns out that the OPF can be non-trivial
only if the participation ratio $\sum_\alpha w_\alpha^2$ is
non-vanishing in the thermodynamic limit. This excludes e.g. the case
of an extensive number of ergodic component where $w_\alpha\sim
\exp(-N\Sigma)$.

Despite its origin in the context of spin glasses, the concept of
overlap, and the corresponding probability function, can be suitably 
to more general glassy systems. A possible  definition in the case of 
structural glasses and some applications are discussed in
\cite{pari-ove,cfp}.

\section{How Aging relates to equilibrium}

Linear response theory allows to relate the OPF of the
canonical probability and the FDR according to the relation \cite{fmpp}:
\begin{eqnarray}
P(q)=\frac{\d x(q)}{\d q}.
\label{equi}
\end{eqnarray}
The argument that leads to (\ref{equi}) is rather formal, and consists
in finding a family of observables, expressed as sum of local
quantities, whose average values relate respectively in equilibrium
and in dynamics to the moments of the OPF and the derivative of FDR
with respect to the correlation. As the dynamical averages should tend
to the equilibrium ones, one deduces the identity between the OPF and the
derivative of the FDR. The arguments runs as follow.  Let us consider
a set of operators which translate the lattice along a direction
$\hat{\imath}$ parallel to one of the coordinate axes:
\begin{equation}
{\cal T}_k^{(p)}(x)=x+\frac{k L}{p} \hat{\imath};\qquad
k=1,\ldots,p-1.
\end{equation}
We denote by ${\cal S}_1$ the set in which the coordinate in the
direction $\hat{\imath}$ takes the values $1,2,\ldots,L/p$. 
We define our family of perturbations to  have the form
\begin{equation}
H_p({\bf S})= \sum_{x\in {\cal S}_1} J_x^{(p)} S_x S_{{\cal
T}_1^{(p)}(x)} \cdots S_{{\cal T}_{p-1}^{(p)}(x)}, 
\label{pert-sr}
\end{equation}
where the couplings $J_x^{(p)}$ are independent, identically
distributed Gaussian variables, with zero mean and variance $E_J
J_x^2=p$. Despite the apparent long range character of the
interactions, the perturbations $H_p$ can be seen as short range
observable in a folded space, and as such they belong to the class
of observables whole long time limit values tend according to our
hypothesis to the equilibrium
ones. Let us consider now a system evolving with Hamiltonian
$H+hH_p$ These perturbations can be seen as short range
perturbation and one can expect their effect to be described by linear
response theory for small $h$. If we consider the expected
values, in off-equilibrium dynamics and at equilibrium of the
perturbation $H_p$, following simple mathematical manipulations
implied by linear response theory and by self-averaging properties
with respect to the variables $J_x^{(p)}$, one finds respectively for
the dynamic and equilibrium averages
\begin{eqnarray}
\langle H_p(t)\rangle_{dyn}&=&-\beta h\int_0^{1} dq p q^{p-1}x(t,q)
\label{uno}
\\
\langle H_p\rangle_{eq}&=&-\beta h(1-\int_0^{1} dq q^{p}P(q))
\label{pio}
\end{eqnarray}
In the ultra-long time limit, $\langle H_p(t)\rangle_{dyn}\to\langle
H_p \rangle_{eq}$ for all $p$. Correspondingly, $x(t,q)\to x(q)$ and
integrating (\ref{uno}) by parts we find the relation (\ref{equi}).
This relation relates the possibility of a persistent non-trivial FDR
in the aging dynamics to to ergodicity breaking in the equilibrium
measure, and as we will discuss, can be taken as the starting point
for an analysis of how aging systems visit configuration space. The
relation can be generalized to other important features which have
been identified theoretically as possible in aging dynamics. One of
this feature is ultrametricity which at equilibrium implies a
hierarchical organization of the ergodic components. In dynamics this
property means that the times $t(C,t_w)$ necessary for the correlation
$C(t,t_w$ to take the value $C$ and defined by the relation
$C(t(C,t_w),t_w)=C$ verify for $C_1<C_2$ the relation:
$t(C_2,t_w)/t(C_1,t_w){\longrightarrow \atop t_w\to\infty}0$, in other
words that decreasing values of the correlation correspond to
increasingly long relaxation times. Reasoning similar to the ones that
lead to (\ref{equi}) allow to conclude that surprisingly in a given
systems static ultrametricity and dynamic ultrametricity either are
both present or both absent. Recently the relation (\ref{equi}) has also been
generalized to local quantities in \cite{loc_pari}.  A more complete
discussion of the relation (\ref{equi}) should qualify the validity of
the use of linear response in presence of ergodicity breaking, which
involves the commutation of the thermodynamic limit and $h\to 0$ in
statics, as well as the long time limit and $h\to 0$ in dynamics. This
property has been called ``stochastic stability'', in reference to the
fact that the commutation of limits is possible whenever the
introduction of weak random perturbations of the kind (\ref{pert-sr})
has not major effects on the statistical distribution of the states
relevant for the equilibrium probability and the long time
off-equilibrium dynamics.  On physical ground one expects stochastic
stability to be valid and linear response theory to have a range of
validity in glassy systems. Violation of this properties should result
in dynamical crossovers for $t\to\infty$ and $h\to 0$ which are in
principle experimentally observable. So if in some systems one could
measure the asymptotic value of the FDR this would lead informations
on the equilibrium free-energy landscape.

\section{Physical picture}

The relation (\ref{equi}), which links off-equilibrium dynamics to the
properties of the free-energy landscape relevant at equilibrium, has
been discussed in the previous section on a purely formal ground: we
have supposed equilibration of 1TO together validity of LRT and
obtained the relation from straight mathematical analysis. The scope
of this section is to discuss its physical origin. In order to
simplify the discussion, we will consider the simplest model of aging
dynamics with a non-trivial FDR. We will suppose that the dynamics
will consists in a short time equilibrium like part and a long time
aging part characterized by a unique age dependent relaxation time
$\tau(t_w)$ which grows and tends to infinity for large $t_w$. Notice
that this is the assumption commonly used to fit aging response data
both in structural glasses \cite{struik} and in spin glasses
\cite{aging}.  In this picture the correlation functions can be
decomposed in a stationary short time part and an aging long time part
as
\begin{equation}
C(t,t_w)=C_{st}(t-t_w)+C_{ag}\left(\frac{t-t_w}{\tau(t_w)}\right) 
\end{equation}
as usual we define $q_{EA}$ as the value that separate the stationary
from the aging part of the correlation so that 
 $C_{st}(t-t_w)=C^*_{st}(t-t_w)-q_{EA}$ is a monotonically
decreasing function equal to $1-q_{EA}$ for $t-t_w=0$ and tending to
zero for $t-t_w\to\infty$, while $C_{ag}(\frac{t-t_w}{\tau(t_w)})$ is
equal to $q_{EA}$ for $\frac{t-t_w}{\tau(t_w)}=0$ and tends to $0$ for
$\frac{t-t_w}{\tau(t_w)}\to\infty$. Correspondingly we will assume that
the FDR will take the value 1 in the stationary domain ($q_{EA}\le
C(t,t_w)\leq 1$) and the constant value $x<1$ in the aging regime
$C(t,t_w)\leq q_{EA}$, so that the linear susceptibility takes the 
form 
\begin{equation}
\chi(t,t_w) = \beta C_{st}(t-t_w)
+ \beta x
C_{ag}\left(\frac{t-t_w}{\tau(t_w)}\right)
\label{chi-p-spin}
\end{equation}.
The relation between correlation and response is often visualized
plotting parametrically $\chi(t,t_w)$ vs. $C(t,t_w)$ for fixed
$t_w$. The present case corresponds to having for large $t_w$ two
straight lines respectively of slope $\beta$ for $q_{EA}< C(t,t_w)\leq
1$ and $\beta x$ for $0\leq C(t,t_w)< q_{EA}$. This form of the
correlation and response is the one found in models of the $p$-spin
model family \cite{cuku}. Numerical simulations of glass-forming systems indicate
that the two regime behavior is a good approximation for the finite
$t_w$ behavior of the response not too close to the estimated value
of $q_{EA}$, although in that case the parameter $x$ depends on $t_w$
\cite{fdr-liq}. In these cases however 1TO are far from equilibrium,
differently from what we suppose here.  

The form (\ref{chi-p-spin}) corresponds to an equilibrium 
distribution where the function $P(q)$ has the simple form
\begin{eqnarray}
P(q)=(1-x) \delta(q-q_{EA})+x \delta(q).
\label{1rsb}
\end{eqnarray}
A straightforward computation shows that the observables $H_p$ of the
previous section are expressed according to (\ref{pio}) as 
\begin{eqnarray}
\langle H_p\rangle= -\beta h(1-q_{EA}^p+xq_{EA}^p). 
\label{pp}
\end{eqnarray}

The equilibrium regime at short times implies that the system spends
its time in regions of the configuration space where it has the time
of approximately equilibrate before relaxing further on a much longer scale. 
This observation can be put at the basis of decomposition 
where the dynamical variables, say the instantaneous values of the 
spins, are written as the sum of a fast and a slow contribution. 
\begin{equation}
S_x(t)=[S_x(t)-m_x(t)]+m_x(t), 
\end{equation}
where $[S_x(t)-m_x(t)]$ is the fast component and the slow component
$m_x(t)$ can be simply defined as
\begin{equation}
m_x(t)=\frac{1}{\tau(t_w)}\int_t^{t+\tau(t_w)}\d u S_x(u).
\end{equation}
The set of the fast variables that correspond to the same slow
variables define metastable regions of configuration space, called
``quasi-states'' that have a life time of the order of $\tau(t_w)$.
This notion can be related to the potential energy landscape notions
of basin or more precisely metabasins \cite{dh} in the inherent
structure picture \cite{sw}, with whom the quasi-states could be
identified.

We can at this point rather naturally define thermodynamic quantities:
to each quasi-state labeled by the values of the slow variables ${\bf
m}=(m_1,...,m_N)$ we can associate a free-energy $f({\bf m})$ and we
can define a ``configurational entropy'' $\Sigma(f)$ as the
logarithmic multiplicity of quasi-states as a function of their
free-energy $f$.

For long but finite times the quasi-states dominating the dynamics
will present a small but extensive free-energy difference with the
lowest state. The parameter $q_{EA}$ has the meaning of self-overlap
within a quasi-state: $q_{EA}=\frac 1 N \sum_x m_x^2(t)$.  The
equilibrium distribution on the other hand concentrates on the ergodic
components, or true states, which have finite free-energy differences
with the ground state and infinite life time. As we have stressed, if
the function $P(q)$ is non trivial the multiplicity of such states
grows less then exponentially with $N$. On the other hand, for the
number of quasi-states with free-energy difference $N\Delta f$ with
the ground state, one can expect: ${\cal N}(f) \sim \exp(N \rho \Delta
f)$.

Let us discuss the effect on the system of a small perturbation and
concentrate for simplicity on the case of $p=1$, corresponding to an
external magnetic field. As discussed first in \cite{mpv}, the effect
of the field on the equilibrium distribution will be twofold. On one
hand within any given ergodic component, with given average
magnetization $M_{av}$, higher weight will be given to configurations
with higher magnetization difference with the average $M-M_{av}$. On
the other, in the selection of the components higher weight will be
given to components with higher average magnetization.  The actual
value the magnetization will take can be then understood as a two step
free-energy minimization process.  The first step is a single
component free-energy minimization.  This is done according to the usual
equilibrium relation within a component $\beta h (M-M_{av})=\frac {1}{N} \beta [\langle
H_1^2 \rangle_{s.c.}-\langle H_1\rangle_{s.c.}^2]= -\beta h
(1-q_{EA})$, where with $\langle \cdot \rangle_{s.c.}$ we have denoted
the average on a single component. The second step is the choice of
the components with the appropriate average magnetization.  The
probability that a given unperturbed component has average
magnetization $M_{av}$ is for small values of $M_{av}$ Gaussian with
zero average and variance $Nq_{EA}$. Taking then into account the
independence of the perturbation from the original Hamiltonian, we
have that the number of quasi-states with free-energy difference
$\Delta f$ with respect to the unperturbed ground state and average
magnetization $M_{av}$ is given by
\begin{equation}
{\cal N}(f,M_{av}) \sim
\exp(N [\rho \Delta f-M_{av}^2/2q_{EA}])
\end{equation}
In presence of the perturbation the actual value of $M_{av}$ will
be that one minimizing $\Delta f-h M_{av}$, respecting the constraint
that the number of states is higher then zero: 
$\rho \Delta f-m^2/2(1-q_{EA})\geq 0$. One finds that the minimization is
achieved in fact for $\rho \Delta f-M_{av}^2/2(1-q^p)=0$ thus finding,
comparing with (\ref{pp})
that $\rho=\beta x$. 

The equivalence (\ref{pio}) tells that the same selection criteria
should be valid asymptotically in off-equilibrium dynamics. The way
the energy distributes among the different degrees of freedom in
dynamics is asymptotically the same as at equilibrium.  Free-energy
not only governs equilibrium, but also dynamics: quasi-states of equal
free-energy are selected asymptotically with equal probability.
Notice that this conclusion do not depend on the observables we have
considered in the analysis, according to our line of reasoning the
same factor $x$ appears in the anomalous response to any perturbation.
Under these circumstances one may wonder how the system can always be
microscopically far from the true equilibrium components as the
off-equilibrium behavior of correlation and response implies. This
relates to the abundance of quasi-states extensive the free-energy
difference $N \Delta f$, which in order to have a non trivial
FDR must be exponentially large. In that case a small $\Delta f$ while
implies small difference for the 1TO from the equilibrium values, 
it implies big microscopic differences between the quasi-states
visited and the ergodic components. 

This considerations, based on time scale separation and vicinity of
1TO to their equilibrium value does not of course directly apply to
structural glasses.  However, based on the analysis of the case of the
$p$-spin model where 1TO block to values different from the
equilibrium ones \cite{fv} one can hypothesize that the previous
quasi-state selection principle could also be valid far from the
ultra-long time limit. This is comforted by numerical simulations
which have reported the existence of well defined FDR very far
from equilibrium which are fairly constant on the aging regime
\cite{fdr-liq} and do not appear to depend on
different quantities which have been studied \cite{barrat}.

In this regime we probe in a region of parameters corresponding to
finite configurational entropy, and as 1TO are strongly out of
equilibrium, no connection can be made between the measurable values
of the FDR and the equilibrium OPF. However, thanks to equiprobability
also in this case the FDR can be related to some static
characterization of the configuration space.  At a given time,
corresponding to a quasi-state free-energy $f$ the system will have
reached with equal probability one of the ${\cal N}(f) \sim \exp(N
\Sigma(f))$ possible quasi-states, where the configurational entropy
$\Sigma(f)$ is an increasing function of $f$. In order to rationalize
the relation (\ref{pp}) one has just to suppose that the dynamics is
dominated by the relaxation of the configurational entropy towards
smaller values, and that a small perturbation in the Hamiltonian do
not change its rate of reduction. In other words, at a given time
$t_w$, the configurational entropy will take the same value
$\Sigma(t_w)$ both in absence and in presence of the perturbation.
Writing then the total free-energy as an unperturbed term
$f$ plus a perturbation term $-\beta h M_{av}$ one can write that
$\Sigma(t_w)=\Sigma(f)-M_{av}^2/2(1-q_{EA})$. Minimizing the total
free-energy $f=f-\beta h M_{av}$ with this constraint one recovers the
formula (\ref{pp}) with the relation
\begin{equation}
\beta
x=\frac{\partial \Sigma}{\partial f}
\label{xxx}
\end{equation}
during aging the effective temperature $1/(\beta x)$ verifies with the
configurational entropy the same relation that the true temperature
verifies at equilibrium with the total thermodynamic entropy. This
kind of relation has been recently used to argue in favor of the the
``Edwards measure'' for lattice gas systems \cite{bkvs} and granular
material under shear \cite{ku-ma}. Let us notice that even if this is
not indicated explicitly in the notation, now the various parameters
($f$, $q_{EA}$ etc.) are slow functions of time.  The equiprobability
hypothesis, if in the ultra-long time limit it comes rather naturally
from 1TO equilibration, is much harder to justify in when these
quantities are out of equilibrium. At present they are not clear the
physical principles that would explain it, although it has been
tentatively related to a chaotic property with respect to the thermal
noise \cite{BBM} according to which two ``clones'', generated doubling
a given system at the time $t_w$ and evolving later with different
realization of the thermal noise, would give rise to divergent
trajectories.

We would like to stress that the analysis we have presented implies
equivalence between the relaxation of field induced perturbations
after removal of the field and the regression of spontaneous
fluctuation \cite{fv} in a way that generalizes the classical Onsager
argument for equilibrium systems \cite{onsager}. Both kind of
conditions in fact imply free-energy minimization with a
configurational entropy constraint, and within the two step relaxation
model we have considered lead to the relation (\ref{chi-p-spin}).

Up to now we have considered as a reference setting a simple quench
from high to low temperature. We can generalize the present dynamical
picture to more complex thermal histories retaining equiprobability of
quasi-states with equal free-energy. We would get in this case a
dynamical picture in agreement with multi-temperature thermodynamics
described in \cite{theo}, where the macroscopic state of a glass is
specified by a suitable number of history dependent effective
temperatures. The model discussed here corresponds to just an
effective temperature in addition to the external one. According to
direct numerical measurement of the FDR in glass-forming models this
corresponds to a good approximation to the situation found in model
liquid systems \cite{fdr-liq}. Good indications in favor of the
present picture come from numerical simulations of glass forming
liquids.  In ref. \cite{scio-tar} the relation (\ref{xxx}) has been
tested with positive answer starting from a direct measure of the FDR
and an estimate of the configurational entropy as the logarithmic
number of inherent structure with given energy.  Other interesting
numerical evidence has been found in the Kob-Andersen kinetic model
in \cite{bkvs} and in the related problem of granular materials under
shear in \cite{ku-ma}. Experimentally there are clear evidence that
violations of FDT are present on long time scales
\cite{israel,cili,ocio}, but if these correspond to effective
temperatures is not clear at the present stage.

\section{Conclusions}

In this paper we have discussed a possible picture on how glassy system visit
configuration space based on the analysis of the fluctuation
dissipation relations during aging. The ultra-long time limit,
possibly relevant for the case of spin glasses, gives us a case where
we can understand in a good detail the meaning of anomalous response
and effective temperatures defined from off-equilibrium fluctuation
dissipation relations. The asymptotic value of anomalous response is
related to the existence of a non trivial equilibrium OPF $P(q)$,
which in turn implies ergodicity breaking in the equilibrium
distribution. The meaning of the equivalence can be found in a
dynamical principle of selection of the quasi-states in the asymptotic
limit: quasi-states with equal free-energy are selected with equal
probability in the dynamical process. Under these circumstances where
the system is macroscopically close to equilibrium, the possibility of
always being microscopically far from equilibrium and continue to age
relates to the abundance of states, which in order to have a non
trivial FDR must be exponentially large as soon in the free-energy
difference $\Delta f$ is finite.

This leads to a dynamical picture that can be generalized to the case
in which the system is macroscopically out of equilibrium, assuming
equiprobability of quasi-states of equal free-energy, and
a rate of decrease of the configurational entropy independent of
possible small perturbations. This hypothesis predicts the existence of
FDR (effective temperatures) related to the growth of the
configurational entropy with quasi-state free-energy and therefore
independent of the particular couple of correlation and response
measured. The experimental verification of this property, and the
clarification of the physical principles leading to it are open
problems for future research.

\section{Acknowledgments}

The present contribution is largely based on papers \cite{fmpp,fv} written
in collaboration respectively with M. Mezard, G. Parisi and L. Peliti
and with M.A. Virasoro, whom I warmly thank.

\end{document}